# Spin-polarized current induced by a local exchange field in a silicene nanoribbon

**Xing-Tao An[1,2], Yan-Yang Zhang[1,3], Jian-Jun Liu[4] and Shu-Shen Li[1]**

[1]SKLSM, Institute of Semiconductors, Chinese Academy of Sciences, P. O. Box 912, Beijing 100083, China

[2]School of Sciences, Hebei University of Science and Technology, Shijiazhuang, Hebei 050018, China

[3]International Center for Quantum Materials, Peking University, Beijing 100871, China

[4]Physics Department, Shijiazhuang University, Shijiazhuang 050035, China

E-mail: anxingtao@semi.ac.cn

**Abstract.** A mechanism to generate a spin-polarized current in a two-terminal zigzag silicene nanoribbon is predicted. As a weak local exchange field that is parallel to the surface of silicene is applied on one of edges of the silicene nanoribbon, a gap is opened in the corresponding gapless edge states but another pair of gapless edge states with opposite spin are still protected by the time-reversal symmetry. Hence, a spin-polarized current can be induced in the gap opened by the local exchange field in this two-terminal system. What is important is that the spin-polarized current can be obtained even in the absence of Rashba spin-orbit coupling and in the case of the very weak exchange filed. That is to say, the mechanism to generate the spin-polarized currents can be easily realized experimentally.We also find that the spin-polarized current is insensitive to weak disorder.

## 1. Introduction

The quantum spin Hall effect[1, 2, 3, 4, 5, 6, 7, 8], a new quantum state of matter with a nontrivial topological property, has been one of the most important topics in condensed matter physics. This novel electronic state has a bulk energy gap separating the valence and conduction bands and a pair of gapless spin-filtered edge states on the sample boundaries. The currents carried by the gapless edge states are immune to nonmagnetic scattering and dissipation due to the protection of time-reversal symmetry. The quantum spin Hall effect was first predicted by Kane and Mele in graphene in which the intrinsic spin-orbit coupling opens a band gap at the Dirac point[1]. However, subsequent works have shown that the intrinsic spin-orbit coupling in graphene is too weak to realize the quantum spin Hall effect under present experimental conditions[9, 10, 11].

Recently, a close relative of graphene, a slightly buckled honeycomb geometry of Si atoms called silicene has been synthesized through epitaxial growth[12, 13]. It has been theoretically shown that silicene is a new massless Dirac Fermion system and the strong spin-orbit coupling in silicene may lead to detectable quantum spin Hall effect[7, 14, 15]. The possibility of dissipationless spin currents along the edges of silicene and the compatibility of silicene with the silicon-based device makes this material particularly attracting for the technological applications in spintronics. Very recently, some theoretical studies have proposed that the pure spin current can take place in a quantum spin Hall system with a point contact or magnetic impurities[16, 17]. In this paper, we theoretically study the electron transport through a silicene nanoribbon in the presence of a exchange field on the upper edge of the nanoribbon, as shown in figure 1. The exchange field parallelled to the nanoribbon plane may arise due to proximity coupling to a ferromagnet such as depositing Fe atoms to the silicene surface or depositing silicene to a ferromagnetic insulating substrate. Due to the local time reversal symmetry breaking by the exchange field, an obvious and experimentally observable spin polarized current can be obtained even for quite weak exchange fields and without Rashba spin-orbit coupling in the system.

## 2. Model and calculation

In the tight-binding representation, the silicene sample with the exchange field can be described by the the following Hamiltonian[15]:

$$
\begin{aligned}
H = & -t \sum_{\langle ij \rangle \alpha} c_{i\alpha}^\dagger c_{j\alpha} + i \frac{\lambda_{SO}}{3\sqrt{3}} \sum_{\langle\langle ij \rangle\rangle \alpha\beta} \nu_{ij} c_{i\alpha}^\dagger \sigma_{\alpha\beta}^z c_{j\beta} \\
& - i \frac{2}{3} \lambda_R \sum_{\langle\langle ij \rangle\rangle \alpha\beta} \mu_{ij} c_{i\alpha}^\dagger (\vec{\sigma} \times \vec{d}_{ij}^0)_{\alpha\beta}^z c_{j\beta} \\
& + M \sum_{i=1,\alpha}^{10} c_{i\alpha}^\dagger \sigma_x c_{i\alpha} + \sum_{i\alpha} c_{i\alpha}^\dagger \varepsilon_i c_{i\alpha},
\end{aligned}
\tag{1}
$$

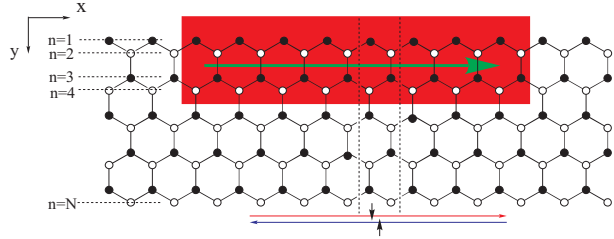

**Figure 1.** (Color online) Schematic diagram of a zigzag silicene nanoribbon with an exchange field (red region) on the upper edge. Silicene consists of a buckled honeycomb lattice of silicon atoms with two sublattices made of A sites (black dots) and B sites (white dots). The green arrow in the red region denotes the direction of the exchange field. The red and blue arrows at the down edges denote propagation directions of the opposite spins in the edge states. The unit cell of the silicene nanoribbon is marked by the dashed lines.

where $c_{i\alpha}^{\dagger}$ creates an electron with spin polarization $\alpha$ at site $i$; $\langle ij \rangle$ and $\langle\langle ij \rangle\rangle$ run over all the nearest and next-nearest neighbor hopping sites, respectively. The first term is the nearest-neighbor hopping with the transfer energy $t = 1.6eV$. The second term describes the effective spin-orbit coupling, where $\vec{\sigma} = (\sigma_x, \sigma_y, \sigma_z)$ is the Pauli matrix of spin and $\nu_{ij}$ is defined as $\nu_{ij} = (\vec{d_i} \times \vec{d_j})/|\vec{d_i} \times \vec{d_j}| = \pm 1$ with $\vec{d_i}$ and $\vec{d_j}$ the two bonds connecting the next-nearest neighbors $\vec{d_{ij}}$. The third term represents the Rashba spin-orbit coupling, where $\mu_{ij} = \pm 1$ for the A (B) site, and $\vec{d_{ij}^0} = \vec{d_{ij}}/|\vec{d_{ij}}|$. The fourth term represents the effect of exchange field with strength $M$, which breaks the local time reversal symmetry on the upper edge of the sample. The local exchange field $M$ may arise due to proximity coupling to a ferromagnet such as depositing Fe atoms to the surface of the upper edge of the silicene nanoribbon or depositing silicene nanoribbon partially to a ferromagnetic insulating substrate. In the fifth term, $\varepsilon_i = E_0 + w_i$ is the on-site energy, where $E_0$ is the energy of the Dirac point, which is set zero as the energy zero point and $w_i$ is the on-site disorder energy uniformly distributed in the range $[-w/2, w/2]$ with disorder strength $w$.

We assume that the the temperature is set to zero, and that two clean and semi-infinite silicene ribbons without exchange fields are employed as left and right leads. The two-terminal conductance of the system can be calculated by the nonequilibrium Green's function method and Landauer-Büttiker formula as

$$G(E) = \frac{e^2}{h}\text{Tr}[\mathbf{\Gamma}_L(E)\mathbf{G}^r(E)\mathbf{\Gamma}_R(E)\mathbf{G}^a(E)], \qquad (2)$$

where $\mathbf{\Gamma}_p(E) = i[\mathbf{\Sigma}_p^r(E) - \mathbf{\Sigma}_p^a(E)]$ is the line-width function and $\mathbf{G}^r(E) = [\mathbf{G}^a(E)]^{\dagger} = 1/[\mathbf{E} - \mathbf{H}_{cen} - \mathbf{\Sigma}_L^r - \mathbf{\Sigma}_R^r]$ is the retarded Green function with the Hamiltonian in the center region $\mathbf{H}_{cen}$[18]. The self-energy $\mathbf{\Sigma}_p^r$ due to the semi-infinite lead-$p$ can be calculated numerically[19].

## 3. Numerical results and analysis

In the following numerical calculations, we use the hopping energy $t$ as the energy unit. The width of the sample and the strength of the effective spin-orbit coupling are chosen as $N = 50$ and $\lambda_{SO} = 0.3t$ in all calculations, respectively. This effective spin-orbit coupling opens a gap at the Dirac points and establishes the quantum spin Hall effect[15, 20, 21].

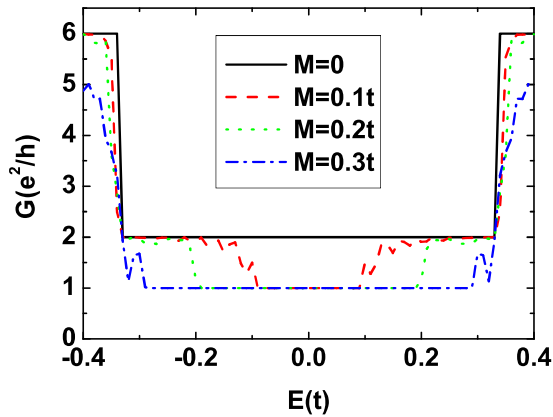

**Figure 2.** (color online) The conductance $G$ of the zigzag silicene nanoribbon vs $E$ for different $M$. The strength of the Rashba spin-orbit coupling is chosen to be $\lambda_R = 0$.

Let us start by describing the influence of the exchange field on the transport properties of the zigzag silicene nanoribbon without disorder, i.e., $w = 0$. In figure 2, the conductance as a function of Fermi energy for different $M$ is studied. In the case of $M = 0$, the conductance $G$ is quantized to a series plateaus at integers (in the unit of $2e^2/h$) due to the transverse sub-bands structure of the system with finite width. In the bulk gap, a quantum conductance plateau with the value $2e^2/h$ appears, coming from the contributions of two pairs of gapless edge states. In the presence of exchange field ($M \neq 0$) on the upper edge of the sample, the conductance plateau $2e^2/h$ is suppressed and evolves into a conductance plateau $e^2/h$ in a wide region in the bulk gap. This is because the gapless edge states on the upper edge are destroyed due to the local time reversal symmetry breaking by the exchange field and the conductance plateau $e^2/h$ is only contributed from the gapless edge states on the lower edge that still protected by time reversal symmetry and persist on the lower edge of the sample. Moreover, the range of the quantized conductance plateau $e^2/h$ is extending with increasing in the magnetization $M$. Especially for $M = 0.3t$, the quantized conductance plateau $2e^2/h$ disappear and the quantized conductance plateau $e^2/h$ extends as far as the corresponding region of the bulk gap opened by the effective spin-orbit coupling. When the energy locates outside of the conductance plateau $e^2/h$, the conductances show slight oscillation between $e^2/h$ and $2e^2/h$ because of the mismatch of interfaces between the leads and the center region.

In order to understand the appearance of the conductance plateau $e^2/h$, we investigate the energy subbands of the silicene nanoribbon for $w = 0$ obtained by solving the lattice model in a strip geometry with open boundary condition in $y$ direction, as shown in figure 3. The gapless chiral edge states can be clearly seen from the band structure of the pristine silicene ribbons ($M = 0$, figure 3(a)). The gapless edge states are robust against small non-magnetic perturbations since they are protected by time reversal symmetry[1]. The electron-hole symmetry is preserved in the gapless edge states, which cross at $ka = \pi$. Due to the spin-up and spin-down states are degenerate, the conductance exhibits a series of plateaus in the unit of $2e^2/h$ for the pristine silicene ribbons, as shown in figure 2. When a local exchange field is applied on the upper edge of the silicene nanoribbon, the corresponding pair of gapless edge states is destroyed and a subgap is opened due to the local time reversal symmetry breaking, as shown in figure 3(b)-(d). Moreover, the magnitude of this subgap opened by the exchange field increases with enhanced $M$, and can be as large as the bulk gap opened by the effective spin-orbit coupling especially for $M = 0.3t$. Another pair of gapless edge states, still protected by time reversal symmetry, persists on the lower edge of the sample. In the surviving gapless edge states, electrons with opposite spin flow in opposite directions along the lower edges of the sample, which lead to quantized conductance plateau $e^2/h$ and the pure spin current. In other words, there remains only right going channels with spin down working for a definite arrangement of bias voltage in this two-terminal system. It is interesting to notice that the subgap of the upper gapless edge states can be opened even in the absence of the Rashba spin-orbit coupling ($\lambda_R = 0$) when a weak exchange field that is parallel to the surface of the silicene nanoribbon is applied at the upper edge of the sample. While the subgap can be opened by a exchange field along the $z$ direction only when there exists the Rashba spin-orbit coupling in the silicene nanoribbon. That is because the Rashba spin-orbit coupling that destroy the electron spin $\sigma_z$ conservation tends to destroy the quantum spin Hall effect[2]. The electron spin $\sigma_z$ conservation can be directly destroyed by the exchange field that is parallel to the silicene nanoribbon plane, so the subgap can be opened in the absence of the Rashba spin-orbit coupling. These suggest that the spin current can be easily achieved experimentally as the local exchange field is parallel to the surface of the silicene nanoribbon.

To test the above arguments about spin polarization in a more direct way, we study the spin-resolved conductance and spin polarization when the local exchange field is applied on the upper edge of the sample. For the sake of simplicity, we assume that in the leads, the effective and Rashba spin-orbit coupling do not exist, i.e., the Hamiltonian of lead-$p$ is simply

$$H_p = -t \sum_{\langle ij \rangle \alpha} c_{i\alpha}^{\dagger} c_{j\alpha}. \tag{3}$$

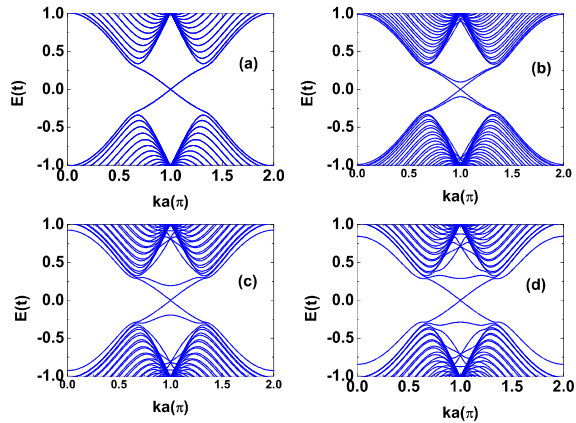

**Figure 3.** (Color online) Calculated energy bands in the zigzag silicene nanoribbon in the presence of the effective spin-orbit coupling for $M = 0$ (a), $M = 0.1t$ (b), $M = 0.2t$ (c), and $M = 0.3t$ (d). The strength of the Rashba spin-orbit coupling is chosen to be $\lambda_R = 0$.

The spin-resolved conductance matrix can be written as

$$G = \begin{pmatrix} G_{\uparrow\uparrow} & G_{\uparrow\downarrow} \\ G_{\downarrow\uparrow} & G_{\downarrow\downarrow} \end{pmatrix}, \tag{4}$$

which can also be calculated by generalized Landauer formula for spin transport. The conductance $G_{\uparrow\uparrow}$ and $G_{\uparrow\downarrow}$ can be obtained when we assume that only spin-up electrons are injected from the left lead into the sample and collected in the right lead. We can also calculate $G_{\downarrow\uparrow}$ and $G_{\downarrow\downarrow}$ in the same way by assuming only spin-down electrons are injected from the left lead. The total conductance $G$ and the spin polarization $P$ in lead-$R$ can be respectively defined as[22, 23]

$$G = G_{\uparrow\uparrow} + G_{\downarrow\uparrow} + G_{\uparrow\downarrow} + G_{\downarrow\downarrow} \tag{5}$$

and

$$P = \frac{G_{\uparrow\uparrow} + G_{\downarrow\uparrow} - G_{\uparrow\downarrow} - G_{\downarrow\downarrow}}{G_{\uparrow\uparrow} + G_{\downarrow\uparrow} + G_{\uparrow\downarrow} + G_{\downarrow\downarrow}}. \tag{6}$$

Figure 4 (a) and (b) show the spin-resolved conductance and spin polarization versus energy $E$ for $M = 0.2t$. In figure 4 (a), the total conductance manifests itself with the plateau value $e^2/h$ because the local exchange field is applied on the upper edge of the sample. Due to the topological nature of the edge states, this plateau is insensitive to the mismatch between the sample and the leads. We can also find that the spin-up and spin-down electrons are not mixed when they transport through the sample, i.e., $G_{\uparrow\downarrow} = G_{\downarrow\uparrow} = 0$. The spin polarization can almost reach 100% in the subgap opened by the local exchange field (see figure 4 (b)) because the spin-down electron can fully transport through the sample while the spin-up electron can hardly transport through the sample in the subgap (see figure 4 (a)). Beyond the gap, due to the conduction band mismatch between the central region and the leads, there are resonant tunneling peaks in spin-resolved conductance and spin polarization beyond the subgap induced by the exchange field.

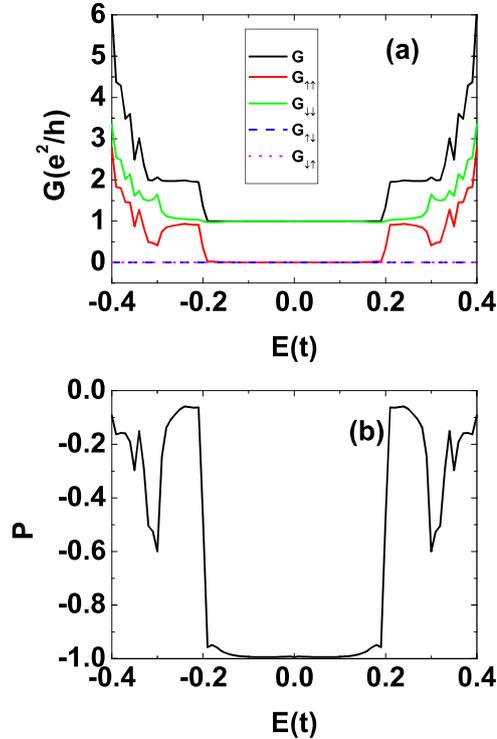

**Figure 4.** (Color online) Spin-resolved conductance (a) and spin polarization (b) vs $E$ for $M = 0.2t$. The parameters $\lambda_{SO} = 0$ and $\lambda_R = 0$ are chosen in leads.

Finally, we examine the effect of the nonmagnetic Anderson disorder on the conductance plateau $e^2/h$ and the spin polarization. The results exhibit that they are stable against the disorder. Figure 5 and figure 6 present the effect of disorder on the conductance and the spin polarization, respectively, where the conductance and the spin polarization are averaged over up to 100 random disorder configurations. Figure 5 (a) shows the conductance $G$ versus the energy $E$ at different $w$ and $M$, and figure 5(b) shows $G$ versus disorder strength $w$ at different $E$ and $M$. The results show that both of the quantum plateaus of $2e^2/h$ (in the case of $M = 0$) and $e^2/h$ (in the case of $M = 0.2t$) are very robust against non-magnetic disorder because of the topological origin of the edge states. The quantum plateaus maintain their quantized values very well even when $w$ reaches $2.0t$. Since the plateau of $e^2/h$ is so robust and stable, the spin-polarized current of the system is insensitive to weak disorder and protected by time-reversal symmetry, as shown in figure 6. Furthermore, when the energy is in close vicinity to the margin value of the subband with subgap opened by the local exchange field (i.e. $E = 0.2t$ or $E = 0.25t$), the values of the conductance are no longer $e^2/h$ and the current is unpolarized as $w = 0$ because the edge state destroyed by the exchange field starts to contribute to the conductance. With the increase of the disorder strength $w$, the conductance plateau $e^2/h$ appear, as shown the blue and green lines in figure 5 (b), because the disorder results in the mobility of the energy band structure[24]. In other words, in this case of energy ($E = 0.2t$ or $E = 0.25t$), the current is unpolarized as $w = 0$, while the spin polarized current can be induced when the disorder strength $w$ increases, as shown the blue and green lines in figure 6 (b). With further increasing of the disorder strength, the conductance gradually reduce to zero and the system eventually enters the insulating regime.

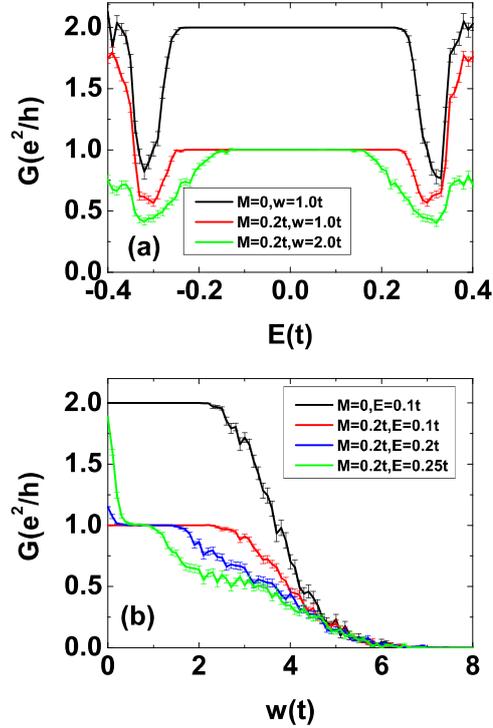

**Figure 5.** The conductance $G$ as a function of the energy $E$ for the various disorder strengths $w$ and various exchange field strengths $M$ (a) and as a function of the disorder strengths $w$ for the various energy $E$ and $M$ (b). The error bars show standard deviation of the conductance for 100 samples.

## 4. Conclusion

In summary, we predict a mechanism to generate a spin-polarized current in a zigzag silicene nanoribbon in the presence of the effective spin-orbit coupling. As a weak exchange field that is parallel to the surface of the silicene is applied on one of edges of the sample, the corresponding gapless edge states is destroyed and a subgap of the subband is opened even in absence of the Rashba spin-orbit coupling. But another pair of gapless edge states with opposite spin are protected by time-reversal symmetry. Therefore, a spin-polarized current with the spin-up and spin-down carriers moving in opposite directions can be observed in this two-terminal system. Furthermore, the spin-polarized current is robust against non-magnetic disorder. The mechanism to generate a spin-polarized current can be easily realized with present technology.

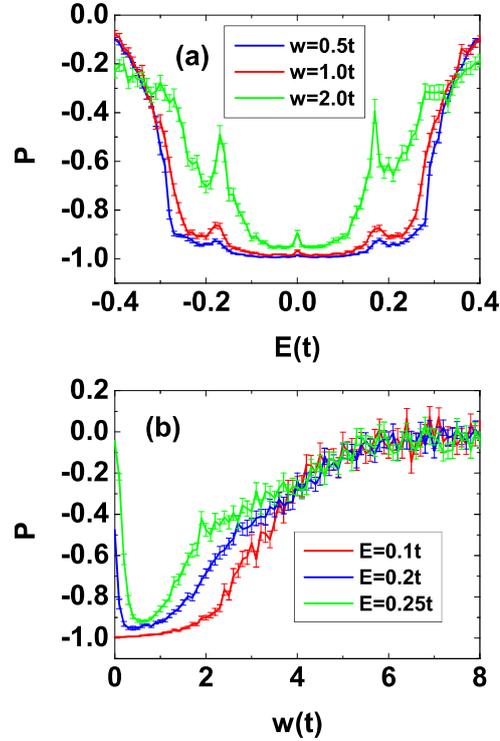

**Figure 6.** The spin polarization $P$ as a function of the energy $E$ for the various disorder strengths $w$ (a) and as a function of the disorder strengths $w$ for the various energy $E$ (b). The exchange field strengths is chosen to be $M = 0.2t$. The error bars show standard deviation of the conductance for 100 samples.

## Acknowledgments

YYZ would like to thank Hua Jiang for helpful discussions. This work was supported by National Natural Science Foundation of China (Grant Nos. 11104059 and 61176089), Hebei province Natural Science Foundation of China (Grant No. A2011208010), and Postdoctoral Science Foundation of China (Grant No. 2012M510523).